\documentclass{iaus}
\usepackage{natbib}
\newcommand{\mnras}{\textit{MNRAS}}
\newcommand{\aap}{\textit{A\&A}}
\newcommand{\aj}{\textit{AJ}}
\newcommand{\apj}{\textit{ApJ}}
\newcommand{\apjl}{\textit{ApJ}}
\newcommand{\araa}{\textit{ARA\&A}}
\newcommand{\nat}{\textit{Nature}}

\newcommand{\oversim}[2]{\protect{\mbox{\lower0.5ex\vbox{%
  \baselineskip=0pt\lineskip=0.2ex
  \ialign{$\mathsurround=0pt #1\hfil##\hfil$\crcr#2\crcr\sim\crcr}}}}}
\newcommand{\simgreat}{\mbox{$\,\mathrel{\mathpalette\oversim>}\,$}} 
\newcommand{\simless} {\mbox{$\,\mathrel{\mathpalette\oversim<}\,$}} 

\title[Pressure-Supported Stellar Systems]{The formation, disruption and properties of pressure-supported stellar
systems and implications for the astrophysics of galaxies}

\author[P. Kroupa]{Pavel Kroupa}

\affiliation{Argelander Institute for Astronomy, Auf dem H\"ugel 71,
D 53121 Bonn, Germany; email: pavel@astro.uni-bonn.de}

\pubyear{2007}
\volume{246}  
\pagerange{}
\date{17th December 2007}
\jname{Dynamical Evolution of Dense Stellar Systems}
\editors{E. Vesperini, M. Giersz, \& A. Sills, eds.}

\begin{document}

\maketitle

\begin{abstract}

Most stars form in dense star clusters deeply embedded in residual
gas. These objects must therefore be seen as the fundamental building
blocks of galaxies. With this contribution some physical processes
that act in the very early and also later dynamical evolution of dense
stellar systems in terms of shaping their later appearance and
properties, and the impact they have on their host galaxies, are
highlighted. Considering dense systems with increasing mass, it turns
out that near $10^6\,M_\odot$ their properties change fundamentally:
stellar populations become complex, a galaxial mass--radius relation
emerges and the median two-body relaxation time becomes longer than a
Hubble time. Intriguingly, only systems with a two-body relaxation
time longer than a Hubble time show weak evidence for dark matter,
whereby dSph galaxies form total outliers.

\keywords{stars: formation, stars: luminosity function, mass function,
globular clusters: general, galaxies: formation}

\end{abstract}

\section{Embedded clusters}
\label{pk_sec:embcl}

{\bf Fragmentation:} The very early stages of cluster evolution on a
scale of a few~pc are dominated by gravitational fragmentation of a
turbulent magnetized contracting molecular cloud core
\citep{Clarkeetal2000,TilleyPudritz2007,MacLowKlessen2004}. The
existing simulations show the formation of contracting filaments which
fragment into denser cloud cores that form sub-clusters of accreting
proto-stars. As soon as the proto-stars emit radiation and outflows of
sufficient energy and momentum to affect the cloud core these
simulations become expensive as radiative transport and deposition of
momentum and mechanical energy by non-isotropic outflows are difficult
to handle with given present computational means
\citep{Stamatellosetal2007,Daleetal2007}.

Observations of the very early stages at times $\simless\,{\rm
few}\,10^5$~yr suggest proto-clusters to have a hierarchical
proto-stellar distribution: a number of sub-clusters with radii
$\simless 0.2$~pc and separated in velocity space are often seen
embedded within a region less than a~pc across
\citep{Testietal2000}. Most of these sub-clusters may merge to form a
more massive embedded cluster
\citep{ScallyClarke2002,FellhauerKroupa2005}.

{\bf Mass segregation:} Whether or not star clusters or sub-clusters
form mass-segregated remains an open issue. Mass segregation by birth
is a natural expectation because proto-stars near the density maximum
of the cluster have more material to accrete. For these, the ambient
gas is at a higher pressure allowing proto-stars to accrete longer
before feedback termination stops further substantial gas inflow
\citep{ZinneckerYorke2007}. Initially mass-segregated sub-clusters
preserve mass segregation upon merging
\citep{McMillanetal2007}. However, in dense proto-clusters (thousands
of stars within less than a pc), the energy equipartition time-scale
between the stars is very short such that mass segregation may
formally occur dynamically within one to a few crossing times
\citep{Kroupa2002Sci}. Currently we cannot say conclusively if mass
segregation is a birth phenomenon (e.g. \citep{Gouliermisetal2004}), or
whether the more massive stars form anywhere throughout the
proto-cluster volume.  Star clusters that have already blown out their
gas are typically mass-segregated (e.g. R136, Orion Nebula Cluster).

Affirming natal mass segregation would impact positively on the notion
that massive stars ($\simgreat 10\,M_\odot$) only form in rich
clusters, and negatively on the suggestion that they can also form in
isolation (for recent work on this topic see
\citet{Lietal2003,ParkerGoodwin2007}).

{\bf Feedback termination:}
The star-formation efficiency (SFE), $\epsilon\equiv M_{\rm
ecl}/$ $(M_{\rm ecl} + M_{\rm gas})$, where $M_{\rm ecl}, M_{\rm gas}$
are the mass in freshly formed stars and residual gas, respectively,
is $0.2 \simless \epsilon$ $\simless 0.4$ \citep{LadaLada2003} implying
that the physics dominating the star-formation process on scales less
than a few~pc is stellar feedback.  Within this volume, the
pre-cluster cloud core contracts under self gravity thereby forming
stars ever more vigorously, until feedback energy suffices to halt the
process ({\it feedback-termination}).

{\bf Dynamical state at feedback termination:} Each proto-star needs
about $t_{\rm ps}\approx10^5$~yr to accumulate about 95~\% of its mass
\citep{WuchterlTscharnuter2003}. The proto-stars form throughout the
pre-cluster volume as the proto-cluster cloud core contracts. The
overall pre-cluster cloud-core contraction until feedback-termination
takes $t_{\rm cl,form}\approx {\rm few} \times (2/\sqrt{G})(M_{\rm
ecl}/\epsilon)^{-{1\over2}}R^{3\over2}$ (a few times the crossing
time), which is about the time over which the cluster forms.  Once a
proto-star condenses out of the hydro-dynamical flow it becomes a
ballistic particle moving in the time-evolving cluster potential.
Because many generations of proto-stars can form over the
cluster-formation time-scale and if the crossing time through the
cluster is a few times shorter than $t_{\rm cl,form}$, then the
assumption may be made that the very young cluster is mostly
virialised when star formation stops and before the removal of the
residual gas. It is noteworthy that $t_{\rm ps}\approx t_{\rm
cl,form}$ for $M_{\rm ecl}/\epsilon \approx 10^{4.5}\,M_\odot$
(proto-star formation time is comparable to the cluster formation
time) which is near the turnover mass in the old-star-cluster mass
function.

A critical parameter is thus the ratio $\tau = t_{\rm cl,form}/t_{\rm
cross}$. If it is less than unity then proto-stars ``freeze out'' of
the gas and cannot virialise in the potential before the residual gas
is removed. Such embedded clusters may be kinematically cold if the
pre-cluster cloud core was contracting, or hot if the pre-cluster
cloud core was pressure confined, because the young stars do not feel
the gas pressure.

In those cases where $\tau > 1$ the embedded cluster is approximately
in virial equilibrium and the pre-gas-expulsion stellar velocity
dispersion in the embedded cluster, $\sigma \approx \sqrt{G\,M_{\rm
ecl}/(\epsilon \, R)}$, may reach $\sigma=40\,$pc/Myr if $M_{\rm ecl}
= 10^{5.5}\,M_\odot$ which is the case for $\epsilon\,R < 1$~pc. This
is easily achieved since the radius of one-Myr old clusters is
$R\approx 0.8$~pc with no dependence on mass. Some observationally
explored cases are discussed by \citet{Kroupa2005}. Notably, using
K-band number counts, \citet{Gutermuthetal2005} appear to find evidence
for expansion after gas removal.  Interestingly, recent Spitzer
results suggest a scaling of $R$ with mass, $M_{\rm ecl}\propto R^2$
\citep{Allenetal2007}, so the question how compact embedded clusters
form and whether there is a mass--radius relation needs further
clarification. I note that such a scaling is obtained for a stellar
population that expands freely with a velocity given by the velocity
dispersion in the embedded cluster. Is the observed scaling then a
result of expansion from a compact birth configuration after gas
expulsion?

There are two broad camps suggesting on the one hand side that
molecular clouds and star clusters form on a free-fall time-scale
\citep{Elmegreen2000,Hartmann2003} and on the other that many free-fall
times are needed \citep{KrumholzTan2007}. The former implies $\tau
\approx 1$ while the latter implies $\tau > 1$.

Thus, currently unclear issues concerning the initialisation of
$N$-body models of embedded clusters is the ratio $\tau = t_{\rm
cl,form}/t_{\rm cross}$, and whether a mass--radius relation exists
for embedded clusters {\it before} the development of HII regions. To
make progress I assume for now that the embedded clusters are in
virial equilibrium at feedback termination ($\tau > 1$) and that they
form highly concentrated with $R\simless 1$~pc independently of mass.

{\bf The mass of the most massive star:} Young clusters show a
well-defined correlation between the mass of the most massive star,
$m_{\rm max}$, in dependence of the stellar mass of the embedded
cluster, $M_{\rm ecl}$, which appears to saturate at $m_{\rm
max}\approx 150\,M_\odot$ \citep{WeidnerKroupa2006}. This may indicate
feedback termination of star formation within the proto-cluster volume
coupled to the most massive stars forming latest, or ``turning-on'' at
the final stage of cluster formation \citep{Elmegreen1983}. The
physical maximum stellar mass near $150\,M_\odot$
\citep{WK04,Figer2005,OeyClarke2005,Koen2006,ZinneckerYorke2007} must
be a result of stellar structure stability, but may be near
$80\,M_\odot$ as predicted by theory if the most massive stars reside
in near-equal component-mass binary systems \citep{KW05}. It may also
be that the calculated stellar masses are significantly overestimated
\citep{Martinsetal2005}.

{\bf The cluster core of massive stars:} Irrespectively of whether the
massive stars ($\simgreat 10\,M_\odot$) form at the cluster centre or
whether they segregate there due to energy equipartition, they
ultimately form a compact sub-population that is dynamically highly
unstable.  Massive stars are ejected from such cores very efficiently
on a core-crossing time-scale, and for example the well-studied Orion
Nebula cluster (ONC) has probably already shot out 70~per cent of its
stars more massive than~$5\,M_\odot$ \citep{PfK2006}. The properties of
O and B runaway stars have been used by \citet{ClarkePringle1992} to
deduce the typical birth configuration of massive stars, finding them
to form in binaries with similar-mass components in compact small-$N$
groups devoid of low-mass stars. Among others, the core of the ONC is
just such a system.

{\bf The star-formation history in a cluster:} The detailed
star-formation history in a cluster contains information about the
events that build-up the cluster. Intriguing is the recent evidence
for some clusters that while the bulk of the stars have ages different
by less than a few~$10^5$~yr, a small fraction of older stars are
often harboured \citep{PallaStahler2000}.  This may be interpreted to
mean that clusters form over about 10~Myr with a final highly
accelerated phase, in support of the notion that turbulence of a
magnetized gas determine the early cloud-contraction phase
\citep{KrumholzTan2007}. 

A different interpretation would be that as a pre-cluster cloud core
contracts on a free-fall time-scale, it traps surrounding field stars
which thereby become formal cluster members: Most clusters form in
regions of a galaxy that has seen previous star formation.  The
velocity dispersion of the previous stellar generation, such as an
expanding OB association, is usually rather low, around a few km/s to
10~km/s. The deepening potential of a newly-contracting pre-cluster
cloud core will be able to capture some of the preceding generation of
stars such that these older stars become formal cluster members
although they did not form in this cluster. \citet{PfK2007} study this
problem for the ONC showing that the reported age spread by
\citet{Pallaetal2007} can be accounted for in this way. This suggests
that the star-formation history of the ONC may in fact not have
started about 10~Myr ago, supporting the argument by
\citet{Elmegreen2000} and \citet{Hartmann2003} that clusters form on a timescale
comparable to the crossing time of the pre-cluster cloud core.

For very massive clusters such as $\omega$~Cen,
\citet{Fellhaueretal2006} show that the potential is sufficiently deep
such that the pre-cluster cloud core may capture the field stars of a
previously existing dwarf galaxy. Up to 30~\% or more of the stars in
$\omega$~Cen may be captured field stars.  This would be able to
explain an age spread of a few~Gyr in the cluster, and is consistent
with the notion that $\omega$~Cen formed in a dwarf galaxy that was
captured by the Milky Way. The attractive aspect of this scenario is
that $\omega$~Cen need not have been located at the center of the
incoming dwarf galaxy as a nucleus, but within its disk, because it
opens a larger range of allowed orbital parameters for the putative
dwarf galaxy moving about the Milky Way. The currently preferred
scenario in which $\omega$~Cen was the nucleus of the dwarf galaxy
implies that the galaxy was completely stripped while falling into the
Milky Way leaving only its nucleus on its current retrograde
orbit. The new scenario allows the dwarf galaxy to be absorbed into
the Bulge of the MW with $\omega$~Cen being stripped from it on its
way in.

{\bf Expulsion of residual gas:}
When the most massive stars are~O stars they destroy the proto-cluster
nebula and quench further star formation by first ionising most of
it. The ionised gas, being now at a temperature near $10^4$~K and in
serious over-pressure, pushes out and escapes the confines of the
cluster volume with the sound speed (near 10~km/s) or faster if the
winds being blown off O~stars with velocities of thousands of km/s
impart sufficient momentum. In reality, this evolution is highly
dynamic and can be described as an explosion (the cluster ``pops''),
and probably occurs non-spherically because the gas seeks low-density
channels in the nebula which then allow the hot gas to escape
\citep{Daleetal2005}.  

If the clusters are more massive than about $10^5\,M_\odot$ such that
the velocity dispersion is larger than the sound speed of the ionised
gas, then the cluster reacts adiabatically because the stars move in a
potential that varies more slowly than the stellar crossing time
through the cluster.  For clusters without O~and massive~B stars,
nebula disruption probably occurs on the cluster-formation time-scale,
$\approx 10^6$~yr, and the evolution is again adiabatic.

\citet{KroupaBoily2002} referred to clusters without O~stars (stellar
mass of the embedded cluster $M_{\rm ecl}\simless 10^{2.5} \,
M_\odot$) as clusters of type~I, those with O~stars but with $10^{2.5}
\simless M_{\rm ecl}/M_\odot\simless 10^{5.5}$ as type~II clusters,
and the very massive clusters ($M_{\rm ecl}\simgreat
10^{5.5}\,M_\odot$) as type~III clusters. A type~IV ``cluster'' may be
added for extremely massive ``clusters'' for which only many
supernovae are able to provide sufficient energy to blow out the
residual gas ($M_{\rm ecl}\simgreat 10^7\,M_\odot$).  This broad
categorisation has easy-to-understand implications for the
star-cluster mass function.

{\it If} clusters pop and which fraction of stars remain in a post-gas
expulsion cluster depend critically on the ratio between the
gas-removal time scale and the cluster crossing time. This ratio thus
defines which clusters succumb to {\it infant mortality}, and which
clusters suffer {\it cluster infant weight loss}.  The well-studied
cases do indicate that the removal of most of the residual gas does
occur within a cluster-dynamical time, $\tau_{\rm gas}/t_{\rm cross}
\simless 1$.  Examples noted \citep{Kroupa2005} are the ONC and R136
in the LMC both having significant super-virial velocity
dispersions. Other examples are the Treasure-Chest cluster and the
very young star-bursting clusters in the massively-interacting
Antennae galaxy which appear to have HII regions expanding at
velocities such that the cluster volume may be evacuated within a
cluster dynamical time.

A simple calculation of the amount of energy deposited by an O~star
into its surrounding cluster-nebula also suggests it to be larger than
the nebula binding energy \citep{Kroupa2005}. This, however, only gives
at best a rough estimate of the rapidity with which gas can be
expelled; an inhomogeneous distribution of gas leads to the gas
removal occurring preferably along channels and asymmetrically, such
that the overall gas-excavation process is highly non uniform and
variable \citep{Daleetal2005}.  \citet{BastianGoodwin2006} note that
many young clusters have a radial-density profile signature expected
if they are expanding rapidly, supporting the notion of fast gas blow
out. \citet{GoodwinBastian2006} and \citet{degrijs_parm07} also find the
dynamical mass-to-light ratia of young clusters to be too large
strongly implying they are in the process of expanding after gas
expulsion.  A more detailed $N$-body study by \citet{BKP08} considers the
change of $\tau_{\rm gas}/t_{\rm cross}$ with the number of OB stars
in clusters of varying mass by comparing the feedback energy in
radiation and winds with the binding energy of the embedded cluster.
These calculations support the sub-division of clusters into~IV types
suggested above.

\citet{Weidneretal2007} attempted to measure infant weight loss by
using a sample of young but exposed Galactic clusters and applying the
maximal-star-mass vs cluster mass relation from above to estimate the
birth mass of these clusters. The uncertainties are large, but the
data firmly suggest that the typical cluster looses at least about
50~per~cent of its stars.

{\bf Mass loss from evolving stars:} An old globular cluster with a
turn-off mass near $0.8\,M_\odot$ will have lost 30~per cent of the
mass that remained in it after gas expulsion due to stellar evolution
\citep{BamgardtMakino2003}. As the mass loss is most rapid during the
earliest times after re-virialisation after gas expulsion, the cluster
expands further during this time. This is nicely seen in the Lagrange
radii of realistic cluster-formation models \citep{KAH}.

\section{Some implications for the astrophysics of galaxies}
\label{pk_sec:impl}

In general, the above have a multitude of implications 
for galactic and stellar astrophysics:

\begin{enumerate}

\item The heaviest-star---star-cluster-mass correlation constrains
feedback models of star cluster formation \citep{Elmegreen1983}.  It
also implies that by adding up all IMFs in all young clusters in a
galaxy, the {\it integrated galaxial initial mass function} (IGIMF) is
steeper than the invariant stellar IMF observed in star clusters with
important implications for the mass--metallicity relation of galaxies
\citep{KW05}

\item \label{pk_point:ML} The deduction that type~II clusters probably
``pop'' implies that young clusters will appear to an observer to be
super-virial, i.e. to have a dynamical mass larger than the luminous
mass \citep{BastianGoodwin2006,degrijs_parm07}.

\item \label{pk_point:thick} It also implies that galactic fields can
be heated, and may also lead to galactic thick-disks and stellar halos
around dwarf galaxies \citep{Kroupa2002}.

\item \label{pk_point:cmf} The variation of the gas expulsion
time-scale among clusters of different type implies that the
star-cluster mass function (CMF) is re-shaped rapidly, on a time-scale
of a few ten~Myr \citep{KroupaBoily2002}.

\item \label{pk_point:popII} Associated with this re-shaping of the
cluster CMF is the natural production of population~II stellar halos
during cosmologically early star-formation bursts
\citep{KroupaBoily2002,ParmGil2007,BKP08}.

\end{enumerate}

Points~\ref{pk_point:ML}--\ref{pk_point:popII} 
are considered in more detail in what follows.

\subsection{Stellar associations, open clusters and moving groups}
\label{pk_ssec:assoc}
As one of the important implications of point~\ref{pk_point:ML}, a
cluster in the age range $1-50$~Myr will have an unphysical $M/L$
ratio because it is out of dynamical equilibrium rather than having an
abnormal stellar IMF \citep{BastianGoodwin2006,degrijs_parm07}.

Another implication is that a Pleiades-like open cluster would have
been born in a very dense ONC-type configuration and that, as it
evolves, a ``moving-group-I'' is established during the first few
dozen~Myr which comprises roughly 2/3rd of the initial stellar
population and is expanding outwards with a velocity dispersion which
is a function of the pre-gas-expulsion configuration \citep{KAH}.
These computations were in fact the first to demonstrate, using
high-precision $N$-body modelling, that the re-distribution of energy
within the cluster during the embedded phase and during the expansion
phase leads to the formation of a substantial remnant cluster despite
the inclusion of all physical effects that are disadvantageous for
this to happen (explosive gas expulsion, low SFE $\epsilon = 0.33$,
Galactic tidal field and mass loss from stellar evolution and an
initial binary-star fraction of 100~per cent).  Thus, expanding OB
associations may be related to star-cluster birth, and many OB
associations ought to have remnant star clusters as nuclei (see also
\citet{Clarketal2005}).

As the cluster expands becoming part of an OB association, the
radiation from its massive stars produce expanding HII regions that
may trigger star formation (e.g. \citet{Gouliermisetal2007}).

A ``moving-group-II'' establishes later as the ``classical'' moving
group made-up of stars which slowly diffuse/evaporate out of the
re-virialised cluster remnant with relative kinetic energy close to
zero. The velocity dispersion of moving group~I is thus comparable to
the pre-gas-expulsion velocity dispersion of the cluster, while moving
group~II has a velocity dispersion close to zero.

\subsection{The velocity dispersion of  galactic-field populations and galactic 
thick disks}
\label{pk_ssec:heat}
Thus, the moving-group-I would be populated by stars that carry the
initial kinematical state of the birth configuration into the field of
a galaxy.  Each generation of star clusters would, according to this
picture, produce overlapping moving-groups-I (and~II), and the overall
velocity dispersion of the new field population can be estimated by
adding in quadrature all expanding populations. This involves an
integral over the embedded-cluster mass function, $\xi_{\rm
ecl}(M_{\rm ecl})$, which describes the distribution of the stellar
mass content of clusters when they are born.  Because the embedded
cluster mass function is known to be a power-law, this integral can be
calculated for a first estimate \citep{Kroupa2002, Kroupa2005}.  The
result is that for reasonable upper cluster mass limits in the
integral, $M_{\rm ecl}\simless10^5\,M_\odot$, the observed
age--velocity dispersion relation of Galactic field stars can be
re-produced.

This theory can thus explain the much debated ``energy deficit'':
namely that the observed kinematical heating of field stars with age
could not, until now, be explained by the diffusion of orbits in the
Galactic disk as a result of scattering on molecular clouds, spiral
arms and the bar \citep{Jenkins1992}. Because the velocity-dispersion
for Galactic-field stars increases with stellar age, this notion can
also be used to map the star-formation history of the Milky-Way disk
by resorting to the observed correlation between the star-formation
rate in a galaxy and the maximum star-cluster mass born in the
population of young clusters \citep{WKL2004}.

An interesting possibility emerges concerning the origin of thick
disks. If the star formation rate was sufficiently high about 11~Gyr
ago, then star clusters in the disk with masses up to
$10^{5.5}\,M_\odot$ would have been born.  If they popped a thick disk
with a velocity dispersion near 40~km/s would result naturally
\citep{Kroupa2002}. The notion for the origin of thick disks appears to
be qualitatively supported by the observations of
\citet{ElmegreenElmegreen2004} who find galactic disks at a redshift
between~0.5 and~2 to show massive star-forming clumps.

\subsection{Structuring the initial cluster mass function}
\label{pk_ssec:mfn}
Another potentially important implication from this theory of the
evolution of young clusters is that {\it if} the
gas-expulsion-time-to-crossing-time ratio and/or the SFE varies with
initial (embedded) cluster mass, then an initially featureless
power-law mass function of embedded clusters will rapidly evolve to
one with peaks, dips and turnovers at cluster masses that characterize
changes in the broad physics involved.

As an example, \citet{KroupaBoily2002} assumed that the function
$M_{\rm icl} = f_{\rm st}\,M_{\rm ecl}$ exists, where $M_{\rm ecl}$ is
as above, $M_{\rm icl}$ is the ``classical initial cluster mass'' and
$f_{\rm st} = f_{\rm st}(M_{\rm ecl})$. The ``classical initial
cluster mass'' is that mass which is inferred by classical $N$-body
computations without gas expulsion (i.e. in effect assuming
$\epsilon=1$, which is however, unphysical). Thus, for example, for
the Pleiades, $M_{\rm cl}\approx 1000\,M_\odot$ at the present time
(age about 100~Myr). A classical initial model would place the initial
cluster mass near $M_{\rm icl}\approx 1500\,M_\odot$ by using standard
$N$-body calculations to quantify the secular evaporation of stars
from an initially bound and virialised ``classical'' cluster
\citep{Portetal2001}. If, however, the SFE was 33~per cent and the
gas-expulsion time-scale was comparable to or shorter than the cluster
dynamical time, then the Pleiades would have been born in a compact
configuration resembling the ONC and with a mass of embedded stars of
$M_{\rm ecl}\approx 4000\,M_\odot$ \citep{KAH}.  Thus, $f_{\rm
st}(4000\,M_\odot) = 0.38 \, (=1500/4000)$.

By postulating that there exist three basic types of embedded
clusters, namely clusters without O~stars (type~I: $M_{\rm
ecl}\simless 10^{2.5}\,M_\odot$, e.g.  Taurus-Auriga pre-main sequence
stellar groups, $\rho$~Oph), clusters with a few O~stars (type~II:
$10^{2.5} \simless M_{\rm ecl}/M_\odot \simless 10^{5.5}$, e.g. the
ONC) and clusters with many O~stars and with a velocity dispersion
comparable to or higher than the sound velocity of ionized gas
(type~III: $M_{\rm ecl}\simgreat 10^{5.5}\,M_\odot$) it can be argued
that $f_{\rm st}\approx 0.5$ for type~I, $f_{\rm st}<0.5$ for type~II
and $f_{\rm st}\approx 0.5$ for type~III. The reason for the high
$f_{\rm st}$ values for types~I and~III is that gas expulsion from
these clusters may be longer than the cluster dynamical time because
there is no sufficient ionizing radiation for type~I clusters, or the
potential well is too deep for the ionized gas to leave (type~III
clusters). Type~II clusters undergo a disruptive evolution and witness
a high ``infant mortality rate'' \citep{LadaLada2003}, therewith being
the pre-cursors of OB associations and Galactic clusters.

Under these conditions and an assumed functional form for $f_{\rm
st}=f_{\rm st}(M_{\rm ecl})$, the power-law embedded cluster mass
function transforms into a cluster mass function with a turnover near
$10^5\,M_\odot$ and a sharp peak near $10^3\,M_\odot$
\citep{KroupaBoily2002}. This form is strongly reminiscent of the
initial globular cluster mass function which is inferred by
e.g. \cite{Vesperini1998,Vesperini2001,ParmGil2005,Baumg1998} to
be required for a match with the evolved cluster mass function that is
seen to have a universal turnover near $10^5\,M_\odot$.

This analytical formulation of the problem has been verified nicely
using $N$-body simulations combined with a realistic treatment of
residual gas expulsion by \citet{BKP08}, who show the Milky-Way
globular cluster mass function to emerge from a power-law
embedded-cluster mass function. \citet{Parmetal2008} expand on this by
studying the effect that different assumptions on the physics of gas
removal have on shaping the star-cluster mass function within about
50~Myr.

The general ansatz that residual gas expulsion plays a dominant role
in early cluster evolution may thus bear the solution to the
long-standing problem that the deduced initial cluster mass function
needs to have this turnover, while the observed mass functions of
young clusters are feature-less power-law distributions.

\subsection{The origin of population~II stellar halos}
\label{pk_ssec:popII}
The above theory implies naturally that a major field-star component
is generated whenever a population of star clusters forms. About
11~Gyr ago, the MW began its assembly by an initial burst of star
formation throughout a volume spanning about 10~kpc in radius. In this
volume, the star formation rate must have reached~$10\,M_\odot$/yr
such that star clusters with masses up to $\approx10^6\,M_\odot$
formed \citep{WKL2004}, probably in a chaotic, turbulent early
interstellar medium. The vast majority of embedded clusters suffered
infant weight loss or mortality, the surviving long-lived clusters
evolving to globular clusters. The so generated field population is
the spheroidal population~II halo, which has the same chemical
properties as the surviving (globular) star clusters, apart from
enrichment effects evident in the most massive clusters. All of these
characteristics emerge naturally in the above model, as pointed out by
\citet{KroupaBoily2002}, by \citet{ParmGil2007} and most recently by
\citet{BKP08}.

\section{Long term, or classical, cluster evolution}
\label{pk_sec:longterm}
The long-term evolution of star clusters that survive infant weight
loss and the mass loss from evolving stars is characterised by three
physical processes: the drive of the self-gravitating system towards
energy equipartition, stellar evolution processes and the heating or
forcing of the system through external tides.  One emphasis of
star-cluster work in this context is on testing stellar-evolution
theory and on the interrelation of stellar astrophysics with stellar
dynamics given that the stellar-evolution and the dynamical-evolution
time-scales are comparable. The reader is directed to other chapters
in this book for further details.

{\bf Tidal tails:} Tidal tails contain the stars evaporating from
long-lived star clusters (the moving~group~II above). Given that
energy equipartition leads to a filtering in energy space of the stars
that escape at a particular time, one expects a gradient in the
stellar mass function progressing along a tidal tail towards the
cluster such that the mass function becomes flatter, i.e. richer in
more massive stars. This effect is difficult to detect, but for
example the long tidal tails found emanating from Pal~5
\citep{Odenkirchenetal2003} may show evidence for this.  As emphasised
by \citet{Odenkirchenetal2003}, tidal tails have another very
interesting use: they probe the gravitational potential of the Milky
Way if the differential motions along the tidal tail can be
measured. They are thus important future tests of gravitational
physics.

{\bf Death:} Nothing lasts forever, and star clusters that survive
initial re-virialisation after residual gas expulsion and mass loss
from stellar evolution ultimately ``die'' after evaporating all member
stars leaving a binary or a long-lived highly hierarchical multiple
system composed of near-equal mass components
\citep{delaFuenteMarcos1997,delaFuenteMarcos1998}. Note that these
need not be stars.  These cluster remnants are interesting, because
they may account for all hierarchical multiple stellar systems in the
Galactic field \citep{GoodwinKroupa2005} with the implication that
they are not a product of star formation, but rather of star-cluster
dynamics.

\section{What is a galaxy?}

Old star clusters, dwarf-spheroidal (dSph) and dwarf-elliptical (dE)
galaxies as well as galactic bulges and giant elliptical (E) galaxies
are all stellar-dynamical systems that are supported by random stellar
motions, i.e. they are pressure-supported. But why is one class of
these pressure supported systems referred to as ``star clusters'',
while the others are ``galaxies''? Is there some fundamental physical
difference between these two classes of systems?

Considering the radius as a function of mass, it becomes apparent that
systems with $M\simless 10^6\,M_\odot$ do not show a mass--radius
relation (MRR) and have $R\approx 4$~pc.  More massive objects,
however, show a well-defined MRR. In fact, \citet{DHK08} find that the
``massive compact objects'' (MCOs), which have $10^6\simless M/M_\odot
\simless 10^8$, lie on the MRR of giant~E galaxies ($\approx
10^{13}\,M_\odot$) down to normal~E galaxies ($10^{11}\,M_\odot$):
$R/{\rm pc} = 10^{-3.15}\,(M/M_\odot)^{0.60\pm0.02}$.  Noteworthy is
that systems with $M\simgreat 10^6\,M_\odot$ also sport complex
stellar populations, while less massive systems have single-age,
single-metallicity populations.  The median two-body relaxation time
is longer than a Hubble time for $M \simgreat 3\times 10^6\,M_\odot$,
and {\it only} for these systems is there evidence for a slight
increase in the dynamical mass-to-light ratio. Intriguingly, $(M/L)_V
\approx 2$ for $M<10^6\,M_\odot$, while $(M/L)_V \approx 5$ for
$M>10^6\,M_\odot$ with a possible decrease for
$M>10^8\,M_\odot$. Finally, the average stellar density maximises at
$M=10^6\,M_\odot$ with about $3\times10^3\,M_\odot/{\rm pc}^3$.

Thus, 

\begin{itemize}

\item the mass $10^6\,M_\odot$ appears to be special,

\item stellar populations become complex above this mass,

\item evidence for dark matter {\it only} appears in systems that have
a median two-body relaxation time longer than a Hubble time,

\item dSph galaxies are the {\it only} stellar-dynamical systems with
$10<(M/L)_V<1000$ and as such are {\it total outliers}. 

\end{itemize}

$M\approx10^6\,M_\odot$ therefore appears to be a characteristic mass
scale such that less-massive objects show characteristics of star
clusters being well-described by Newtonian dynamics, while more
massive objects show behaviour more typical of galaxies. Defining a
galaxy as a stellar-dynamical object which has a median two-body
relaxation time longer than a Hubble time, i.e. essentially a system
with a smooth potential, may be an objective and useful way to define
a ``galaxy'' \citep{Kr98}.

Why {\it only smooth} systems show evidence for dark matter remains at
best a striking coincidence, at worst it may be symptomatic of a
problem in understanding dynamics in such systems.

\end{document}